\documentclass[12pt,a4paper,dvips]{article}
\usepackage{a4p}
\usepackage{cite,mcite}
\usepackage{graphicx}
\usepackage{physics}
\usepackage{l3_title,ifthen}
\usepackage{Lep}
\usepackage{amsmath}
\usepackage{dcolumn}
\journalname{Phys. Lett. B}

\date{February 09, 2000}
\preprint{2000-044}
\Lep{1}
\newlength{\capindent}
\newlength{\capwidth}
\newlength{\figwidth}
\newcommand{\icaption}[2][!*!,!]{\hspace*{\capindent}%
  \begin{minipage}{\capwidth}
    \ifthenelse{\equal{#1}{!*!,!}}%
      {\caption{#2}}%
      {\caption[#1]{#2}}
  \end{minipage}}
\newcommand{\FE}{F^\gamma_{eff}}
\newcommand{\FF}{F^\gamma_2}
\newcommand{\FL}{F^\gamma_L}

\def\ZPC { Z. Phys.}
\newcolumntype{-}{D{-}{~\mbox{--}~}{4}}
\newcolumntype{/}{D{/}{~\pm~}{2}}
\newcolumntype{+}{D{+}{~\pm~}{9}}
\newcolumntype{a}{D{a}{~\pm~}{11}}

 {\end{list}}
  
\begin{document}
\begin{titlepage}
\title{Measurement of the Photon \\
Structure Function at High {\boldmath$Q^2$} at LEP}
\author{The L3 Collaboration}
%
% The abstract
%
\begin{abstract}
The structure functions of real and virtual photons are derived
from cross section measurements of the reaction
$\rm e^+e^-\rightarrow e^+e^-  + \hbox{ hadrons}$ at LEP.
The reaction is studied at $\rts \simeq 91 \GeV $  with the L3
detector. One of the final state electrons is detected at a large angle 
relative to the beam direction, leading to $Q^2$ values between
40 \GeV$^2$ and 500 \GeV$^2$.
The other final state electron is either undetected or it is detected at
a four-momentum transfer squared $P^2$ between 1 \GeV$^2$ and 8 \GeV$^2$.
These measurements are compared with predictions of the Quark Parton
Model and other QCD based models.
\end{abstract}
%
% Adds "To be submitted to ..." or "Submitted to ..." if relevant
%
 \submitted
\end{titlepage}
%
%%%%%%%%%%%%%%%%%%%%%%%%%%%%%%%%%%%%%%%%%%%%%%%%%%%%%%%%%%%%%%%%%%%%%%%%%%%%%%%
%
% Introduction
%
%%%%%%%%%%%%%%%%%%%%%%%%%%%%%%%%%%%%%%%%%%%%%%%%%%%%%%%%%%%%%%%%%%%%%%%%%%%%%%%
%
\section{Introduction}
Deep-inelastic electron scattering on a photon target is interesting
because of its potential to test predictions of quantum chromodynamics (QCD)
\cite{ref:Witten}. In the $Q^2$ domain of the present investigation,
$40 - 500 \GeV ^2$, the photon structure function $\FF$ is dominated
by the pointlike contribution, which can be calculated by perturbative QCD,
and which rises logarithmically with $Q^2$. In addition, there is a 
non-perturbative hadronic contribution, which is usually derived from a
Vector Dominance (VDM) ansatz\cite{ref:Vogt}.

\begin{figure}[htbp]
  \begin{center}
  \includegraphics[width=0.45\textheight]{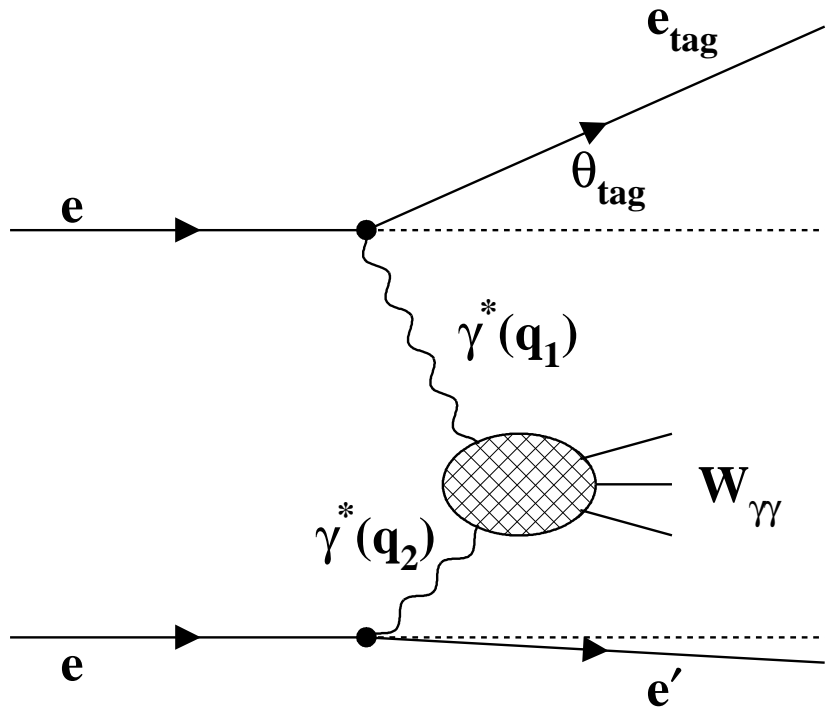}
  \caption{ Diagram of a two-photon interaction 
            $\rm e^+e^-\rightarrow e^+e^- + \hbox{ hadrons}$;
            $q_1$ and $q_2$ are the four-momentum vectors of the probe and target virtual
            photons, $W_{\gamma\gamma}=\sqrt{(q_1+q_2)^2}$ 
            is the two-photon centre-of-mass energy.
            The electron from the lower vertex is either undetected
            or observed at a small angle.
           \label{fig:dis}}
\end{center}
\end{figure}

The deep-inelastic scattering of electrons on a photon target
is studied by measuring the two-photon reaction
$ \rm e^+ e^-\rightarrow e^+ e^- + \hbox{ hadrons}$,
tagged by the detection of one of the final-state electrons\footnote
{Electron stands for electron or positron throughout this paper.}
at a large four-momentum transfer $q_1$ (Figure \ref{fig:dis}).
The other electron usually escapes at a small scattering angle, thereby
ensuring that the target photon is nearly on shell, 
$q_2^2\approx 0$ ({\it single-tag events}). 
Sometimes it is detected at small angles, with a small four-momentum transfer $q_2$
({\it double-tag events}).
The two virtual photons, 
with virtuality $ Q^2\equiv -q_1^2$ and $ P^2\equiv -q_2^2$,
 are
referred to as the ``probe'' and ``target'' photons, respectively.

The differential cross section for the process
$  \rm e^+e^-\rightarrow e^+e^-\gamma^*\gamma^*\rightarrow e^+e^-X$ 
is given in Ref.\cite{ref:Budnev}.
After integration over the azimuthal angles of the outgoing electrons,
it depends on four independent helicity cross sections $ \sigma_{ ab}$
for virtual photon-photon collisions, where $a,b = L,T$ indicate longitudinal
and transverse polarizations of the probe and target photons in the
$\gamma^*\gamma^*$ centre-of-mass.
In the double-tag configuration investigated in this paper,
$\sigma_{LL}$ is negligible in the Quark Parton Model
(QPM), while in the single-tag configuration
$\sigma_{TL}$ and $\sigma_{LL}$ are both negligible \cite{ref:Budnev}. 

Introducing the hadronic mass squared $ W_{\gamma\gamma}^2=(q_1+q_2)^2$,
the energy and polar angle of the first scattered electron
$E_{tag}$ and $\theta_{tag}$, and the energy of the incoming electrons
$E_{beam}$, the Bjorken scaling variables $x$ and $y$ are given by:
\begin{equation}
\label{eq:xy}
 x = {Q^2\over Q^2+P^2+W_{\gamma\gamma}^2},
\enskip\enskip\enskip y = 1-{E_{tag}\over E_{beam}}\cos^2{\theta_{tag}\over 2}.
\end{equation}
For a real photon target ($ P^2=0$) the differential cross section
of the deep-inelastic scattering reaction
$\rm  e\gamma\rightarrow e  + \hbox{ hadrons}$ can  be written as
\begin{equation}
\label{eq:dsigxy}
{d\sigma\over dxdy} = {8\pi\alpha^2\over Q^4} E_{beam}E_\gamma\{[1+(1-y)^2]
\FF-y^2\FL\},
\end{equation}
where $E_\gamma$ is the target photon energy and
\begin{equation}
\label{eq:f2}
\FF(x,Q^2) = {Q^2\over 4\pi^2\alpha} [\sigma_{TT}(x,Q^2)+\sigma_{LT}(x,Q^2)]
\quad\hbox{ and }\quad\enskip
\FL(x,Q^2) = {Q^2\over 4\pi^2\alpha}\sigma_{LT}(x,Q^2).
\end{equation}
By convention $\FF/\alpha$ is measured, where $\alpha $ is the fine-structure constant.
In order to obtain the $\rm e^+e^-$ cross section a convolution
with the flux function of the target photon is necessary.
For the single-tag condition the $\FL$ dependent term in 
equation (\ref{eq:dsigxy}) is less than 5\% in the QPM\footnote
{ For $y<0.7$ the factor $y^2/ (1+(1-y)^2)$ is on average of
the order of 0.2 and  the QPM predicts $ \FL/\FF$ to be smaller than 0.25.
 This leads to $[y^2/ (1+(1-y)^2)]\FL/\FF < 0.05$.}.
In the double-tag measurement $\sigma_{TL}$ cannot be neglected; consequently
one can only measure an effective structure function
\begin{equation}
\label{eq:feff}
\FE={Q^2\over 4\pi^2\alpha} (\sigma_{TT} + \sigma_{LT} + \sigma_{TL} + 
\sigma_{LL}).
\end{equation}
\par 
The photon structure function has been measured at various 
$Q^2$ values at the PETRA, PEP and TRISTAN accelerators \cite{ref:EXPTS} 
and at LEP \cite{ref:OPAL ,ref:DELPHI , ref:L3 , ref:ALEPH }.
The structure function for double tag was first measured by the
PLUTO collaboration\cite{ref:PLUTODT}. 
Here we report on the analysis of events with $Q^2$ in the range 
$40-500 \GeV ^2$ and $P^2$ either close to zero or in the range $1-8 \GeV ^2$.
The data were collected by the L3 detector during the years $1991-1995$  at
$ \sqrt{s}=89-92 \GeV $ with a total integrated luminosity of 120 $\rm pb^{-1}$.
%
%%%%%%%%%%%%%%%%%%%%%%%%%%%%%%%%%%%%%%%%%%%%%%%%%%%%%%%%%%%%%%%%%%%%%%%%%%%%%%%
% 
%  Kinematic fitting
%
%%%%%%%%%%%%%%%%%%%%%%%%%%%%%%%%%%%%%%%%%%%%%%%%%%%%%%%%%%%%%%%%%%%%%%%%%%%%%%%
%
\section{Kinematic fitting} 

A new feature in the present work is a fit to two-photon kinematics 
imposed on each event. Inputs to the fit are the measurements of the 
kinematic variables of the hadrons and of the scattered electrons, 
 constrained to four-momentum conservation.
The single- and double-tag cases are treated differently.

\par
In the single-tag case, the event plane is taken to be the scattering
plane, defined by the directions of the tagged electron and the beam.
In the double-tag case, the event plane is defined by the direction of the vector
sum of the momenta of the tagged electrons and by the beam direction.
The components of the  momentum vectors in the event plane are labelled
$p_z$ and $ p_{in}$, parallel and perpendicular to the beam respectively;
$ p_{out}$ is the component perpendicular to this plane.
The positive $z$ direction is defined in the direction of the tagged
electron. 
From the energy and momenta of the hadronic system, $ (E^h, p^h)$,
of the scattered positron, $ (E^+, p^+)$, and of the scattered electron,
$ (E^-, p^-)$, three energy and momentum constraints are defined: 
\begin{eqnarray}
 C_1 = E^-+E^+ +E^h -2E_{beam}+ 2K_{rad},\enskip
 C_2 = p_{z}^- + p_{z}^+ + p_z^h,\enskip
 C_3 = p^-_{in} + p^+_{in} + p_{in}^h. 
\end{eqnarray}
The term $K_{rad}$ 
describes the average energy carried away
by unobserved initial state radiative photons (ISR). 
In the fit, the following expression is minimized with respect to three
components of the four-momentum vector  of the hadronic system:
\begin{equation}
\label{chisq}
\chi^2 = \sum_{i,j=1}^3 V^{-1}_{i,j}\Delta p_i\Delta p_j + \sum_{k=1}^3
 W_k C_k^2.
\end{equation}
Four energy and momentum differences are defined as:
\begin{eqnarray}
\label{deltap}
\Delta p_1 = E^h - \sum_l E_l^h,\enskip
\Delta p_2 = p_z^h - \sum_l p_{z,l}^h,\enskip
\Delta p_3 = p_{in}^h - \sum_l p_{in,l}^h,\enskip
\Delta p_4 =  p_{out}^h - \sum_l p_{out,l}^h,
\end{eqnarray}
where the index $l$ runs over all detected hadrons and $E^h$, $p_z^h$,
$p_{in}^h$ are the fit quantities.
The $3\times 3$ error matrix, $V$, of the hadronic energy measurement is
calculated from the individual cluster energy and momentum measurements 
by taking into account errors in energy measurements and
uncertainties in the hadron directions. 
The constraints $C_k$ are applied with constant weight factors $W_k$.
These weight factors reflect the accuracy of the measurements
on the energies of the tagged electrons and the spread introduced by
radiative corrections.
The distribution of $\Delta p_4$ is not used in the fit but to compare
the hadronic energy resolution between data and Monte Carlo.

\par
In the single-tag case the unseen electron is assumed to have 
zero scattering angle. 
As one of the
electron momenta is unknown, the constraints $C_1$ and $C_2$ are
combined into:
\begin{equation}
\label{eq:dpl}
 C_{12} = {E^h+p_z^h\over 2} - E_{beam}+ K_{rad} + {E_{tag}+p_{z,tag}\over 2}.
\end{equation}

\par
Most two-photon Monte Carlo programs have no provision for ISR.
The value of $K_{rad}$, estimated by a program of Berends, Daverveldt and
Kleiss\cite{ref:BKD}, is of the order of $1.0-1.5 \GeV $ for a beam energy
of $45.6 \GeV$. It increases logarithmically with the tagging angle.
In fits to the data such losses are taken into account by using 
$K_{rad}=1.5 \GeV $ for the large-angle electron 
 and $K_{rad}=1.0 \GeV $ for the small-angle electron in
double-tag events.
\par
Before the fit the mean values of $\Delta p_i$, with $i=1-3$, are adjusted to zero
by scaling the hadronic energies and momenta. The scale factor accounts
for the average reduction in hadron energy from the loss of particles.
The hadron energy resolution is estimated from the widths of the
constraint distributions. The scaling procedure improves the measurement
of the visible hadronic mass.
\par
The kinematic fit , applied to each event, determines the fitted value
of the hadronic invariant mass and, for single-tag events,
the energy of the unobserved electron, $ E_{mis}$:
\begin{equation}
\label{eq:invmass}
 W_{\gamma\gamma,fit}^2= E^{h2}-p_{in}^{h2}-p_z^{h2},\enskip\phantom{000}
       E_{mis} = E_{beam} - {0.5(W_{\gamma\gamma,fit}^2+Q^2)\over 2E_{beam}
       - (E_{tag}+p_{z,tag})}.
\end{equation}

\begin{figure}[htbp]
  \begin{center}
  \includegraphics[height=0.7\textheight]{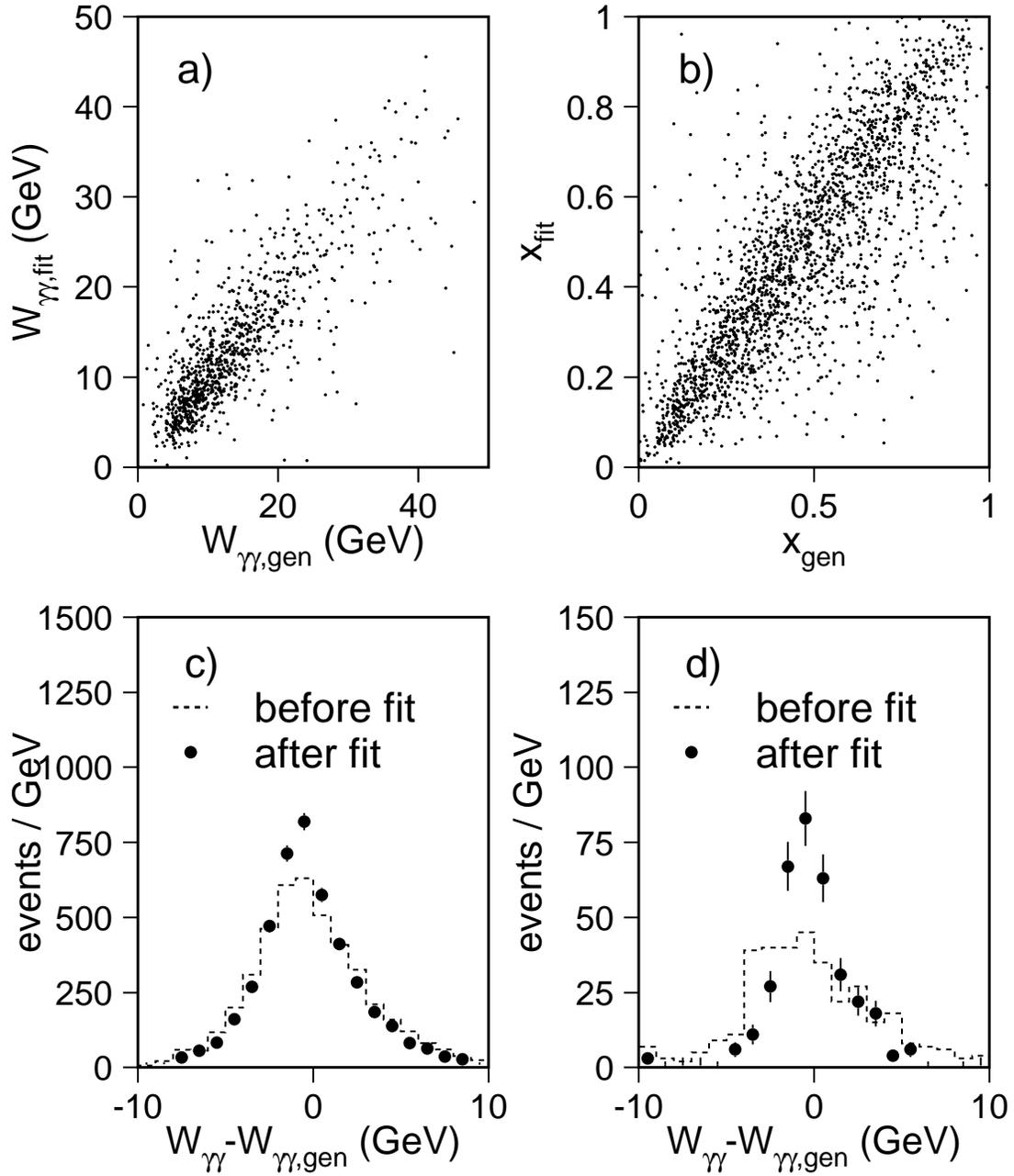}
  \caption{a) Correlation between the generated value of $W_{\gamma\gamma}$
              and the measured value after the kinematic fit. In all cases the
              JAMVG Monte Carlo has been used.
           b) Correlation between the generated value of $x$ and the measured value
           after the kinematic fit.
           c) The hadronic mass resolution before and after the
           fit for single-tag events. 
           d) The hadronic mass resolution before and after the
           fit for double-tag events. 
           \label{fig:xmatrix-bst}}
  \end{center}
\end{figure}

The correlation between the generated values of  $W_{\gamma\gamma}$ and
$x$ and their fitted values for single-tag \qqbar {} events 
is displayed in Figure \ref{fig:xmatrix-bst}; their relation is
approximately linear.
The kinematical fit improves the hadronic mass resolution by about 12\% in
single-tag events and by 30\% in double-tag events.
After the kinematical fit, the one sigma mass resolution is about $3.5 \GeV $
for single-tag and $2.7 \GeV $ for double-tag events; the resolution in $x$
is 0.13 for single-tag and 0.10 for double-tag events. The resolution
is constant in $x$ and $Q ^2$ in the ranges covered by the experiment.
%
%%%%%%%%%%%%%%%%%%%%%%%%%%%%%%%%%%%%%%%%%%%%%%%%%%%%%%%%%%%%%%%%%%%%%%%%%%%%%%%
% 
%  Event selection
%
%%%%%%%%%%%%%%%%%%%%%%%%%%%%%%%%%%%%%%%%%%%%%%%%%%%%%%%%%%%%%%%%%%%%%%%%%%%%%%%
%
\section{Event selection}
A detailed description of the L3 detector and its performance is given in Ref.
\cite{ref:L3det}. 
The analysis in this paper is based on the central tracking system,
the high resolution electromagnetic calorimeters and the hadron calorimeters.
The electron scattered at high $Q^2$ is observed in the endcap electromagnetic
calorimeter at a polar angle $\theta$ between 200 and 700 mrad with respect
to the direction of one of the beams. The other electron remains either
undetected or it is observed in one of the small-angle electromagnetic
calorimeters in a fiducial region 29 mrad $ \le \theta \le 67$ mrad.
The hadronic energy and momentum and the visible mass are derived from the
energy clusters in the electromagnetic and hadron calorimeters.
Energy clusters in the small-angle calorimeters are also included
if their energy is less than $18 \GeV $.
\par
The single-tag events are accepted by two independent triggers, a charged
particle track trigger \cite{ref:L3tec} and a calorimetric energy
trigger\cite{ref:L3en}.
% which includes the signal from the scattered electron.
The average efficiency of the combination of both triggers, deduced from
the data, is $95\pm 1$\%.
It decreases to 85\% at the lowest accepted visible hadronic mass
of $3 \GeV $. For the double-tag events the energy trigger also accepts a
small-angle electron in coincidence with one charged particle.
This yields a trigger efficiency larger than 98\% for double-tag events.
\par
The selection of the process $\rm e^+e^-\rightarrow e^+e^- + \hbox{hadrons}$
has been guided by fully simulated event samples from several
$\gamma\gamma$ Monte Carlo generators. 
The JAMVG\cite{ref:Vermaseren} $\gamma\gamma$ generator is based on an exact 
calculation of the multiperipheral diagrams. The 
$\gamma\gamma \rightarrow \uubar$ and $\gamma\gamma \rightarrow \ccbar$
 channels are generated separately   
with $ m_{\mathrm{u}} = 0.325 \GeV $ and $ m_\mathrm{c}=1.6 \GeV $.
The contributions from d- and s-quarks are taken into account by a 
multiplicative factor 9/8 to the $\rm u\bar u$ cross section.
PHOJET\cite{ref:PHOJET} is an event generator, within the 
Dual Parton Model framework.
The photon is considered as a superposition of a ``bare photon'' and virtual
hadronic states. 
%PHOJET was designed for quasi-real photon interactions.
To separate soft from hard processes, a transverse momentum cutoff
at $2.5 \GeV $ is applied to all the partons of the pointlike interactions
\cite{ref:Field}.
TWOGAM\cite{ref:TWOGAM} generates three different $\gamma\gamma$ processes
separately: the QPM, soft hadronic VDM and the QCD resolved photon 
contribution with a transverse momentum cutoff at $2.5 \GeV $.
\par
The dominant background processes are $\rm e^+e^-\rightarrow\hbox{hadrons}$,
simulated by JETSET\cite{ref:Sjostrand}, $\rm e^+e^-\rightarrow\tau^+\tau^-$
simulated by KORALZ\cite{ref:Jadach} and 
$\rm e^+e^-\rightarrow e^+e^-\tau^+\tau^-$, simulated by JAMVG.
\par
\begin{figure}[htbp]
  \begin{center}
  \includegraphics[width=0.7\textheight]{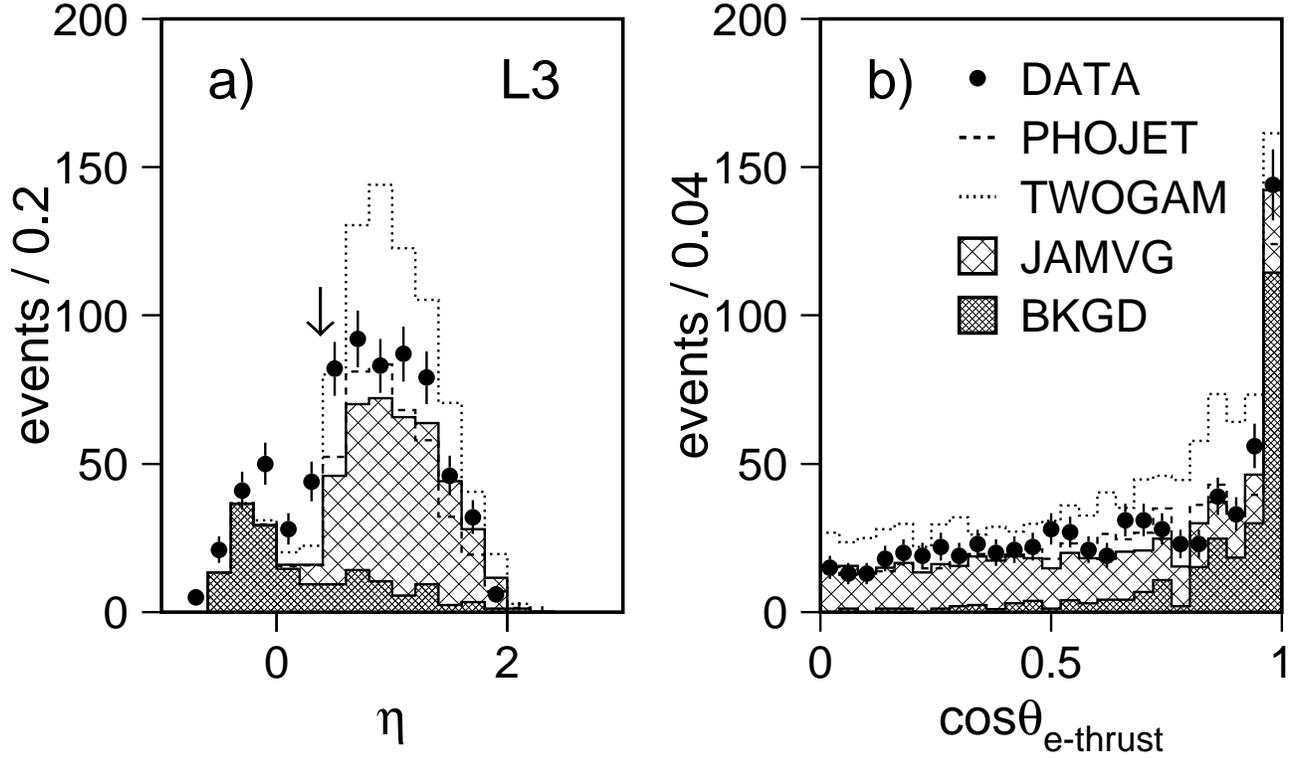}
  \caption{a) The distribution of the event rapidity, $ \eta$, for single-tag
              events; the arrow indicates the cut at $ \eta=0.4$.
           b) The cosine of the angle between the tagging electron and
              the thrust direction in the $\gamma\gamma$ centre-of-mass for
              single-tag events.
              The cut on $\cos\theta_{\mathrm{e-thrust}}$ is mass dependent,
              see the text.
           All selection cuts are fulfilled, except the one on the plotted
           variable. The Monte Carlo distributions, 
           JAMVG, PHOJET, TWOGAM and the background
           from annihilation and two-photon $\tau -$pair production,
           are normalized to the integrated luminosity of the data.
           \label{fig:bstsel-7}}
\end{center}
\end{figure}

Single-tag hadronic events are selected as follows:
\begin{enumerate}
\item[$\bullet$] The scattered electron candidate is an electromagnetic
            cluster with an energy greater than 30\% of the beam energy.
\item[$\bullet$] At least three tracks are seen in the central tracking system.
\item[$\bullet$] The visible hadronic mass $W_{\gamma\gamma,vis}$ is greater
                 than $3 \GeV $.
\item[$\bullet$] The hadronic transverse momentum component,
                 $ p_{out}^h$, is less than $5 \GeV $ and \\
                 $ p_{in}^h \cdot  p_{in}^{tag} < 0$,
                 where $p^{tag}$ stands for the momentum
                 of the tagged electron.
\item[$\bullet$] The event rapidity, $ \eta$, must be greater than 0.4,
      see Figure \ref{fig:bstsel-7}a; $ \eta$ is defined by
      \begin{equation}
      \label{eq:srapbgo}
      \eta = {1\over 2} \ln {{\sum_i (E_i+p_{z,i})\over\sum_i (E_i -p_{z,i})}},
      \end{equation}
      where $i$ runs over the calorimetric clusters, including that of the
      scattered electron.
      This requirement suppresses the annihilation reaction, where
      the total longitudinal momentum of the event is close to zero.
      In the two-photon reaction it can be of the order of $40 \GeV $
      because of the momentum of the unobserved electron.
\item[$\bullet$] The $\chi^2$ probability in the kinematic fit is greater
     than $10 ^ {-5}$.
\item[$\bullet$] The momentum of the unobserved electron must be greater
      than $26 \GeV $.
\item[$\bullet$] 
      A thrust $T$ and a thrust axis $\hat n$ are defined by maximising the 
      linear sum of projected hadronic cluster momenta along
      $\hat n$\cite{ref:Farhi}
      \begin{equation}
      \label{eq:thrust}
      T = {Max \sum_i |\vec p^{\hskip.2em\relax h,*}_i \cdot \hat n|
         \over \sum_i |\vec p^{\hskip.2em\relax h,*}_i|}.
      \end{equation}
      Here $\vec p^{\hskip.2em\relax h,*}_i$ refers to the hadronic cluster
      momentum in the $\gamma\gamma$ centre-of-mass.
      A variable cut is applied on the maximum allowed value of the 
      cosine of the electron-thrust angle in the $\gamma\gamma$ centre-of-mass
      if the hadronic mass is greater than $10 \GeV $;
      its value is 0.92, 0.80, 0.70, 0.60 
      for masses smaller than 20, 30, 40, 50 \GeV , respectively.
      This cut discriminates against the annihilation reaction, 
      in particular it removes $\rm  e^+e^-\rightarrow b\bar b$ events, where the b-quark
      decays semileptonically. In that case the electrons are emitted
      predominantly in the thrust direction, 
       whereas no such correlation is present in
      the two-photon reaction (see Figure \ref{fig:bstsel-7}b). 
\end{enumerate}
\par
Double-tag hadronic events are selected in a similar way to
single-tag events. The small-angle scattered electron candidate
must have an energy greater than 40\% of the beam energy, and  no
cut is applied to the event rapidity.
\par
Table 1 shows the numbers of events selected
in the range $40 \GeV ^2 \le Q^2 \le 500 \GeV ^2 $ and 
$3 \GeV  \le W_{\gamma\gamma,fit} \le 50 \GeV $.
The largest contamination in the 
single-tag sample comes from two-photon $\tau$-pair production.
The influence of the annihilation background in single-tag events is 
checked by comparing the data taken at the Z peak with the off-peak
data (25\% of the sample) where it is about a factor three smaller
than on the peak. Within the limited accuracy the off-peak sample is
consistent with the total selected sample. 
%
%%%%%%%%%%%%%%%%%%%%%%%%%%%%%%%%%%%%%%%%%%%%%%%%%%%%%%%%%%%%%%%%%%%%%%%%%%%%%%%
% 
%  Comparison of Data with Models
%
%%%%%%%%%%%%%%%%%%%%%%%%%%%%%%%%%%%%%%%%%%%%%%%%%%%%%%%%%%%%%%%%%%%%%%%%%%%%%%%
%
\section{Comparison of Data with Models}
\begin{figure}[htbp]
  \begin{center}
  \includegraphics[height=0.7\textheight]{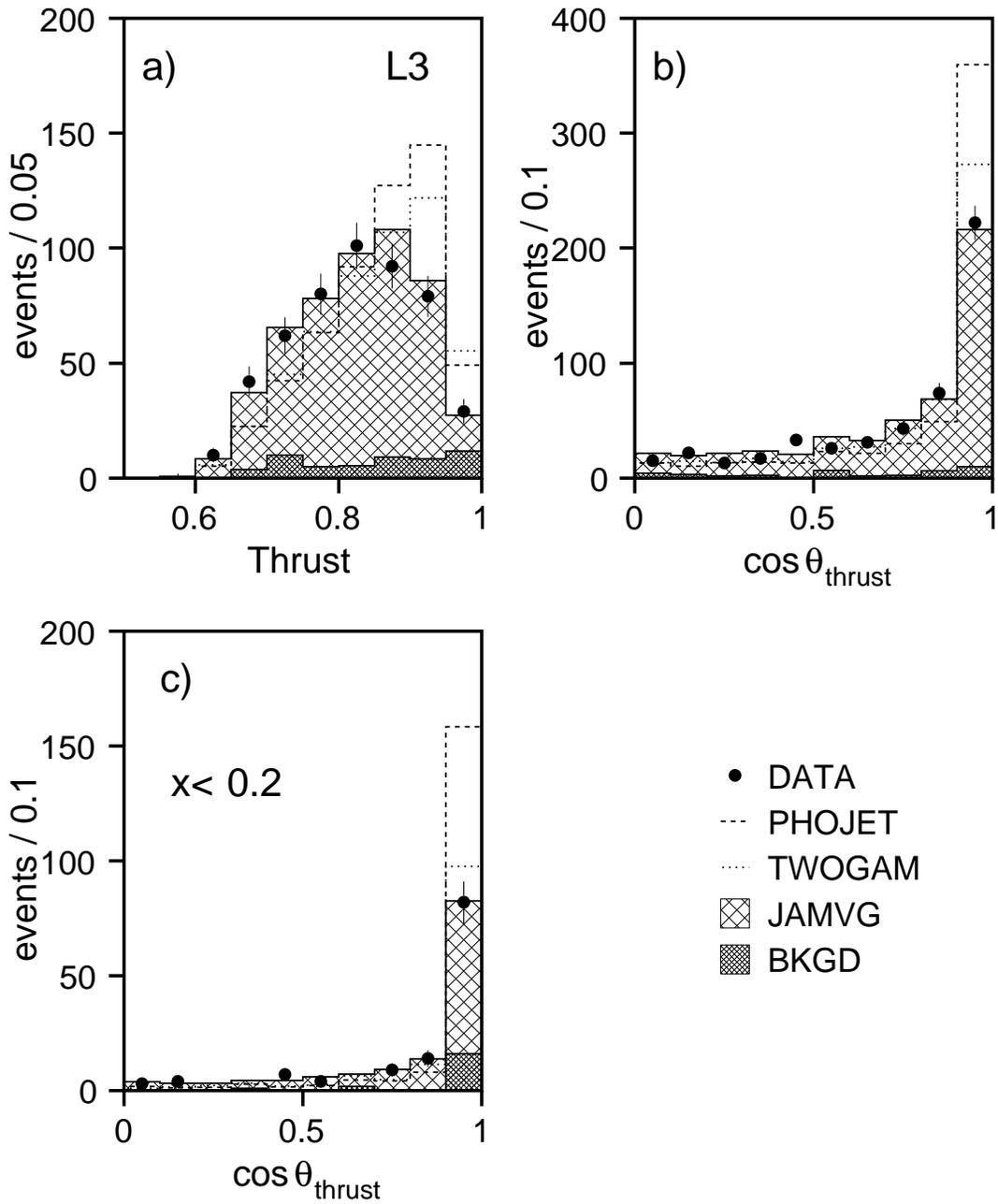}
  \caption{The distributions of a)  thrust, b) $\rm \cos{\theta_{thrust}}$ and 
         c) $\rm \cos{\theta_{thrust}}$ for events
           with $x_{fit}<0.2$ in the $\gamma\gamma$ centre-of-mass frame for
           single-tag.
         The JAMVG, PHOJET and TWOGAM contributions are summed with the 
         background and scaled to have the same number of simulated events
         as in the data.
           \label{fig:tcpc-bst}}
  \end{center}
\end{figure}

\begin{figure}[htbp]
  \begin{center}
  \includegraphics[height=0.7\textheight]{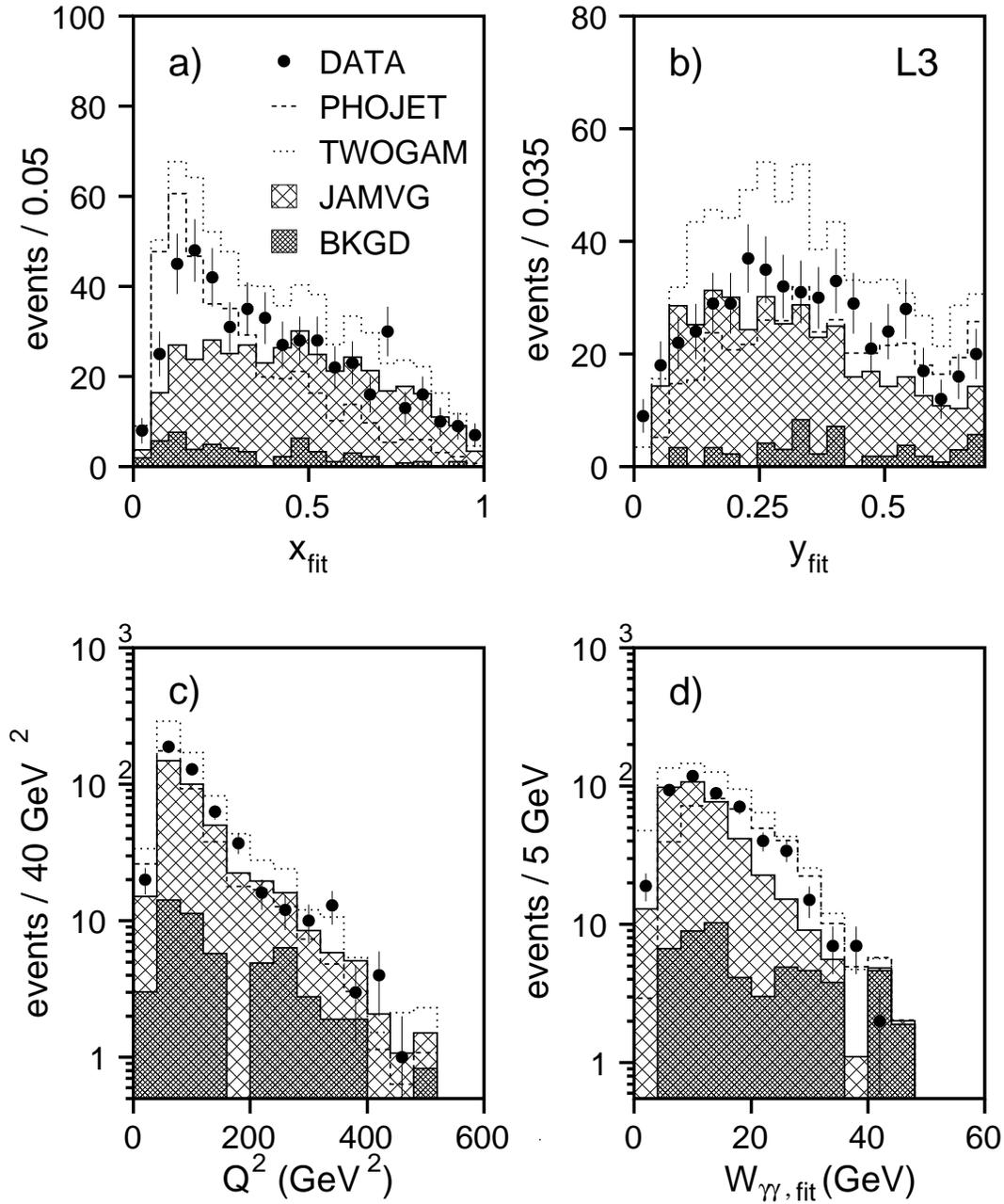}
  \caption{The a) $ x_{fit}$ , b) $ y_{fit}$,
           c) $ Q^2$ , d) $ W_{\gamma\gamma,fit}$, distributions for single-tag events.
           The data are compared to the JAMVG, PHOJET and TWOGAM models.
           The predictions include the estimated background distributions.
           The Monte Carlo distributions are normalized to the integrated luminosity
           of the data.
           \label{fig:xyqw-bst}}
  \end{center}
\end{figure}

\begin{figure}[htbp]
  \begin{center}
  \includegraphics[height=0.7\textheight]{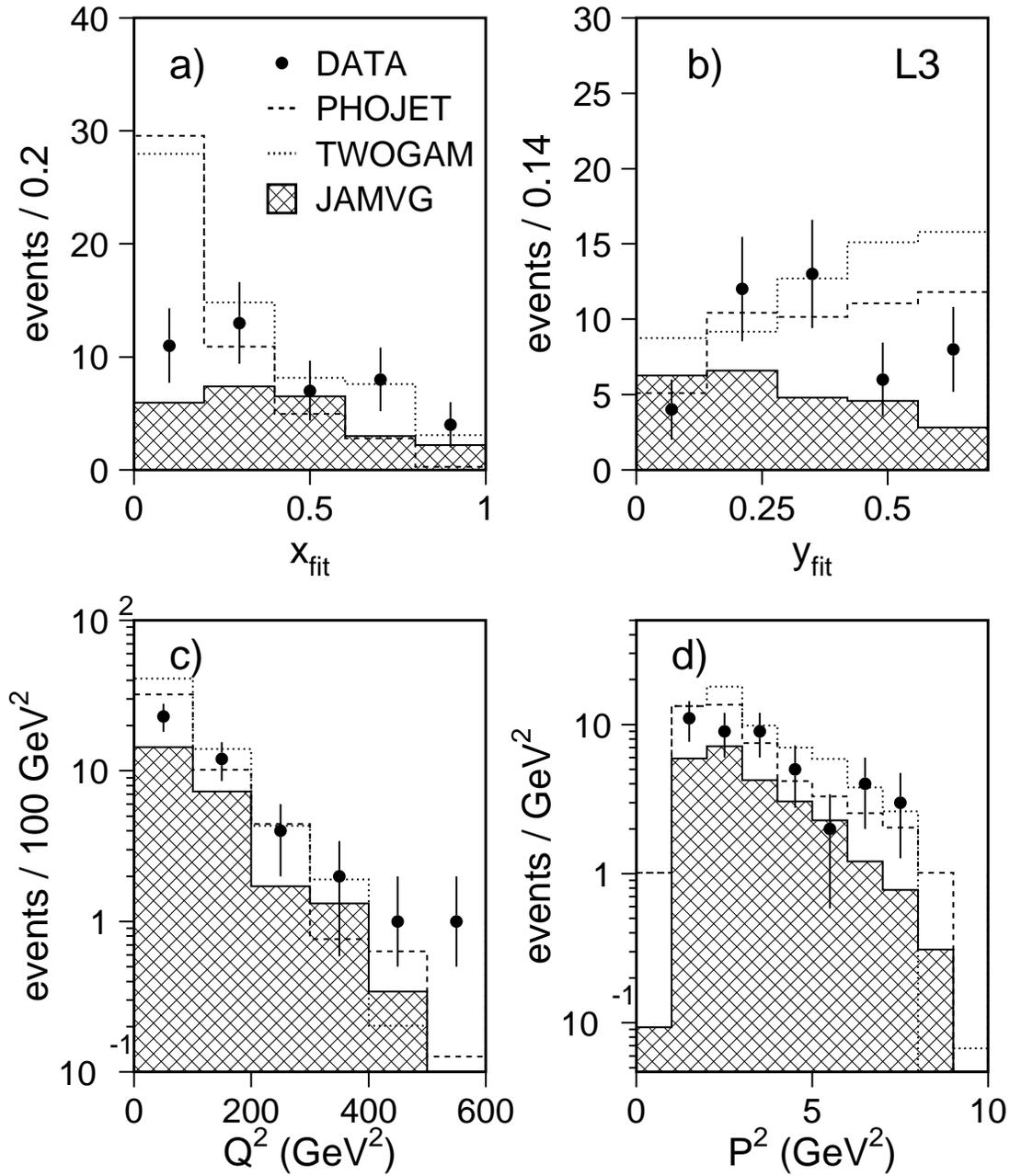}
  \caption{The a)   $ x_{fit}$, b)$ y_{fit}$ , c) $Q^2$, d) $ P^2$
            distributions for double-tag events.
           The data are compared to the JAMVG, PHOJET and TWOGAM predictions.
           Backgrounds are estimated to be negligible.
           The Monte Carlo distributions are normalized to the integrated luminosity of
           the data.
           \label{fig:xyqp-bdt}}
  \end{center}
\end{figure}

The data are compared to Monte Carlo expectations in Table 1
and in Figures \ref{fig:tcpc-bst}, \ref{fig:xyqw-bst} and
\ref{fig:xyqp-bdt}.
In Figure \ref{fig:tcpc-bst}  kinematic properties in the $\gamma\gamma$ 
centre-of-mass of the selected single-tag events are compared in shape
to Monte Carlo predictions. For this purpose the Monte Carlo distributions
are normalized to the same number of events as the data.
In \ref{fig:tcpc-bst}a) the thrust exhibits a wide distribution, which is
adequately described by JAMVG (QPM) and less well by the PHOJET and TWOGAM
models. In \ref{fig:tcpc-bst}b) and \ref{fig:tcpc-bst}c) (for $x<0.2$)
the distribution of the cosine of the polar angle of the thrust direction
with respect to the $\gamma\gamma$ axis ($\theta_{thrust}$) shows forward
peaking in good agreement with the JAMVG simulation. A diffractive forward
peak, predicted by PHOJET and TWOGAM, is not observed in the data.
\par
It should be noted that the compatibility of the QPM angular distributions
with the data does not exclude the contribution from  QCD processes,
such as photon-gluon fusion and $\rm \gamma q\rightarrow gq$ which
have similar angular distributions\cite{ref:Brodsky}, but 
indicates a low contribution of VDM and diffractive processes.
\par
Figures \ref{fig:xyqw-bst} a)$-$d) show the distributions of $x_{fit}$, $ y_{fit}$, $ Q^2$ and
$ W_{\gamma\gamma,fit}$,  together with Monte Carlo
background estimations, for the selected single-tag events, in comparison
with Monte Carlo predictions normalised to the data luminosity.
Although the number of events expected by JAMVG is too small, the $ Q^2$ distribution follows the predicted shape.
The hadronic mass $ W_{\gamma\gamma,fit}$ presents an excess 
of data over JAMVG predictions at masses above $20 \GeV $.
The $ x_{fit}$ distribution is limited by the maximum mass observed,
$ W_{\gamma\gamma,fit}\approx 50 \GeV $, and by the minimum mass of
$ W_{\gamma\gamma,fit}=3 \GeV $ and by the restricted $Q^2$ range from
40 \GeV$^2$ to 500 \GeV$^2$.
The data agree with the prediction 
of JAMVG for $x_{fit}>0.5$. For lower $x_{fit}$ values
the QPM alone is insufficient to reproduce the data. 
The excess of events over the JAMVG prediction at $ x_{fit}<0.2$ and
$ y_{fit}>0.3$ is compatible with the existence of a QCD contribution
in the measured $Q^2$ range.
TWOGAM predicts too many events at all $x_{fit}$
values.  PHOJET agrees with the data at low
$x_{fit}$, but not at high $x_{fit}$ because this program suppresses also
the QPM diagram with the cutoff on the quark transverse momentum
($p_t > 2.5 \GeV$).
Also the \ccbar {} production is suppressed in PHOJET, whereas
the JAMVG program predicts a substantial charm contribution in
both single- and double-tag events.
\par
The double-tag events have a negligible background from Z-decay and
$\tau$-pairs.  Figures \ref{fig:xyqp-bdt} a)$-$d) show the distributions
for $x_{fit}$, $y_{fit}$, $Q^2$ and $P^2$,  together with the Monte Carlo
expectations.
Also here the QPM model implemented in JAMVG predicts too few events,
but  the kinematic distributions, such as
the thrust, the thrust angular distributions and the $ p_t$ distribution
of the energy clusters, are well reproduced.
PHOJET and TWOGAM expectations exceed the data at low $x_{fit}$ values, thus 
indicating that the  QCD contributions are overestimated.
%
%%%%%%%%%%%%%%%%%%%%%%%%%%%%%%%%%%%%%%%%%%%%%%%%%%%%%%%%%%%%%%%%%%%%%%%%%%%%%%%
% 
%  Photon structure functions
%
%%%%%%%%%%%%%%%%%%%%%%%%%%%%%%%%%%%%%%%%%%%%%%%%%%%%%%%%%%%%%%%%%%%%%%%%%%%%%%%
%
\section{Photon structure functions}
\subsection{Single tag}
The JAMVG generator reproduces the shape of the kinematic distributions. Therefore
it provides the basis for calculating the acceptance and it can be
used to unfold the $ x_{fit}$ distributions to distributions in true
values of $x$.
After subtraction of background, the single-tag data are divided in five
$ x_{fit}$ bins, and unfolded into $ x$ bins with the SVD method\cite{ref:SVD}.
The unfolding procedure is considered to be satisfactory if the $ x_{fit}$
distribution, calculated from the unfolded result,
reproduces the measured $ x_{fit}$ distribution within its statistical errors.
The acceptance has a maximum of about 60\% at $x = 0.4$ and minima of about
40\% at both ends of the $x$ interval.
\par
\begin{figure}[htbp]
  \begin{center}
  \includegraphics[height=0.7\textheight]{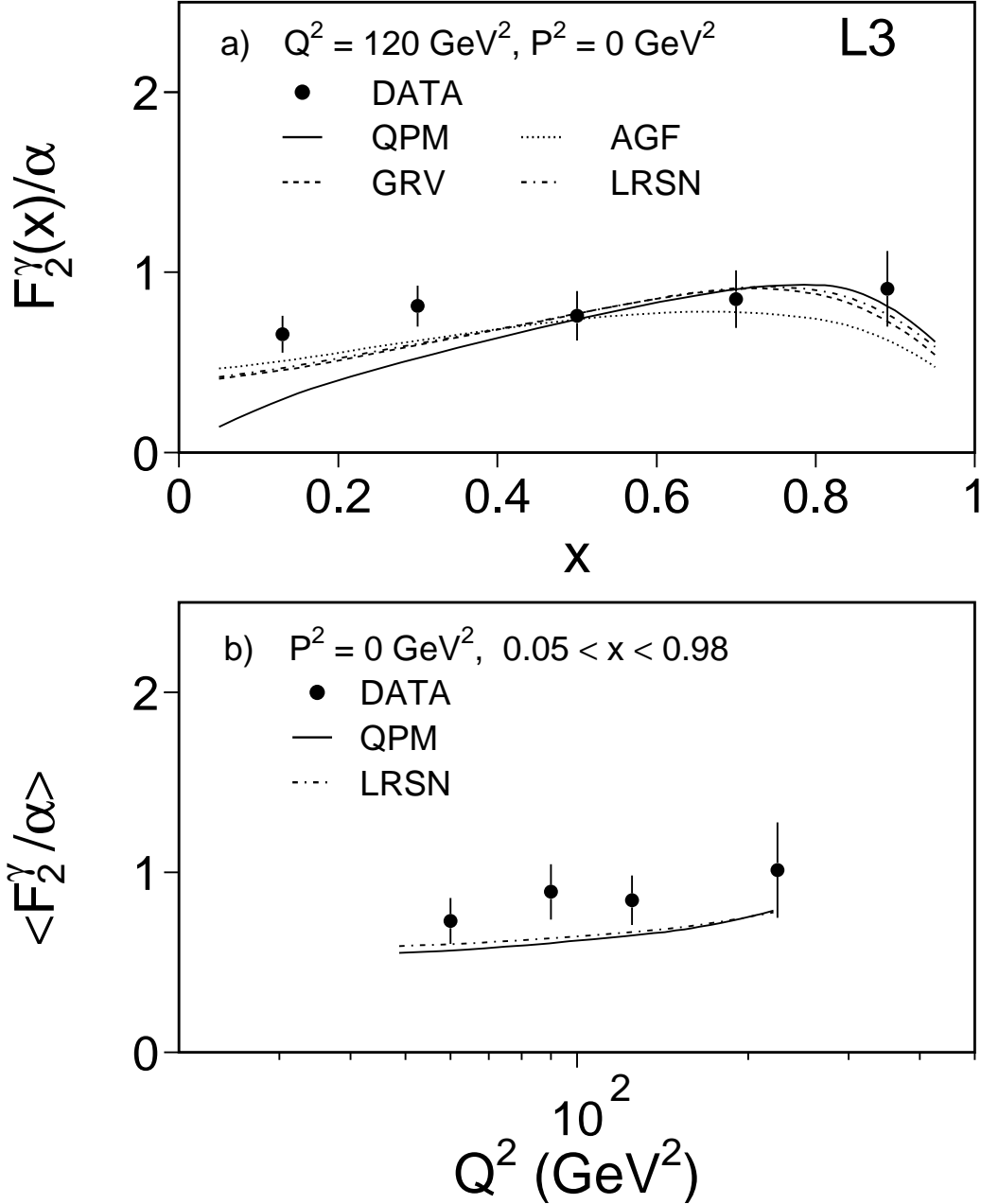}
  \caption{a) The structure function $\FF/\alpha$ for real photons at
        $ Q^2=120$ \GeV$^2$ compared with the QPM calculation
        and the QCD calculations GRV, AGF and LRSN described in the text.
        b) Dependence on $ Q^2$ of $\FF/\alpha$ averaged over $x=0.05-0.98$
        for single-tag data, compared with the QPM indicated
        by a full line and with the LRSN calculation described in the text.
        The errors are statistical and systematic added in quadrature.
      \label{fig:plot-fsingle}}
  \end{center}
\end{figure}

The value of $\FF/\alpha$ is obtained by comparing the experimental $x$ 
distribution to the generated one.
The ratios, given in Table 2, are applied to the QPM 
analytical expression for $\FF(x)/\alpha$, obtained from the 
$\sigma_{TT}$ and $\sigma_{LT}$ formul\ae, given in appendix E of Ref.\cite{ref:Budnev},
 calculated for every $\langle x\rangle$ value
at the average $ Q^2 $ of the data distribution, $ Q^2=120$ \GeV$^2$,
 and at $ P^2=0$. The  QPM estimates  that 
 the structure function, calculated at the
the mean value of the target photon virtuality 
  $ P^2=0.014 \GeV^2$,   is only 0.2 \%  smaller
than the value expected for a real photon,
this effect is therefore neglected.
 The result is shown in Figure \ref{fig:plot-fsingle}
and in Table 2. The correlation between the statistical errors,
introduced by the unfolding procedure, is small.
The correlation matrix is also given in Table 2.
\par
Estimates of systematic uncertainties from various sources on each data point
are summarized in Table 3, averaged over the entire $x$ and $ Q^2$ range.
The misidentification of hadrons as tagged electrons
is small due to the requirement of  a substantial tag energy;
the corresponding uncertainty on the tag selection is estimated by comparing
  the results of the different generators.
The model dependence in the acceptance calculation and in the
selection efficiency as function of $ x_{fit}$ is also evaluated  
 from the Monte Carlo generators.
JAMVG includes only the multiperipheral two-photon diagram.
The contributions from the bremsstrahlung, annihilation and conversion diagrams
missing in JAMVG have been estimated for the single-tag configuration with
the program DIAG36 \cite{ref:DIAG36}.
They are small, and consistent with zero with a 2.4\% uncertainty.
The systematic uncertainties in the event selection as defined in Section 3 have been
estimated by varying the cut parameters.
An estimate of the effect of radiative corrections on the
cross section has been made by Laenen and Schuler\cite{ref:Laenen}.
In the region of interest they find a reduction of the cross section of the order
of $2.6-3.8$\%. The average  is taken to be the systematic uncertainty.
No radiative correction is applied to the data.
The annihilation background has been estimated by varying the cut on the
variable $\eta$, shown in Figure \ref{fig:bstsel-7}a).
By unfolding the data with different generators and by using a bin-by-bin
correction a mean spread per bin of 5.3\% is observed and it is assigned to the
systematic uncertainty. 
The total systematic uncertainty is estimated to be 9.6\%, 
almost independent of $x$.

The resulting values of $\FF(x)/\alpha$ at $ Q^2 =120$ \GeV$^2$ can be compared
with various calculations.
In Figure \ref{fig:plot-fsingle}a) a comparison is
shown with the QPM predictions  \cite{ref:Budnev}, 
with the QCD models  GRV
\cite{ref:GRV} and AGF \cite{ref:AGF}  and with the LRSN
next-to-leading order (NLO) QCD calculations 
\cite{ref:LRSN}.
As already observed by comparing the $x_{fit}$ distribution of data with 
Monte Carlo, at high $x$ the data are well described by the QPM,
but QCD contributions are necessary for $ x< 0.5$.
The existing QCD models predict similar values, but  all  are below
the experimental data at low $x$.
\par
The value of $\FF(x)/\alpha$ at $\langle x\rangle =0.13$ can be compared to our measurements 
at lower values of $ Q^2$\cite{ref:L3}.
The ln$ Q^2$ evolution obtained there extrapolates to a value
$\FF /\alpha = 0.63\pm 0.13 \stat \pm 0.13 \sys$ at  $Q^2 =120$ \GeV$^2$,
 in agreement with the measured value. 
Figure \ref{fig:plot-fsingle}b) and Table 4 show the $ Q^2$-dependence of
$\FF/\alpha$ for the data, the QPM model \cite{ref:Budnev} and the LRSN
QCD calculation  after integration over the range
$x=0.05-0.98$. The data are higher than the model predictions.
 The QPM prediction depends on the assumed values of the
quark masses. If the quark masses are reduced by 30\%, the average value of 
$\FF/\alpha$ increases by about 5\%.
The $Q^2$ evolution is compatible with a ln $Q^2$ rise.
For comparison with current other data \cite{ref:EXPTS ,ref:OPAL ,ref:DELPHI ,
ref:L3 , ref:ALEPH}, the results in the ranges $x=0.3-0.8$ and
$x=0.1-0.6$ are given in Table 4.
We also give there $\langle\FF/\alpha\rangle$ at 
$Q^2=120$ \GeV$^2$, averaged over the $Q^2$ range of the experiment.
%
%%%%%%%%%%%%%%%%%%%%%%%%%%%%%%%%%%%%%%%%%%%%%%%%%%%%%%%%%%%%%%%%%%%%%%%%%%%%%%%
% 
%  Double tag
%
%%%%%%%%%%%%%%%%%%%%%%%%%%%%%%%%%%%%%%%%%%%%%%%%%%%%%%%%%%%%%%%%%%%%%%%%%%%%%%%
%
\subsection{Double tag} 
The procedure to derive the effective structure function $\FE(x)$
at average values $ Q^2=120$ \GeV$^2$ and  $ P^2=3.7$ \GeV$^2$
is the same as that for single-tag data.
The resulting values of $\FE(x)/\alpha$ are given in Table 5. 
The total systematic uncertainty is of the same order as that of the single-tag case;
the selection is simpler and backgrounds are negligible, but the
Monte Carlo and unfolding errors are larger. The total error is dominated by
 the statistical error.
\par
\begin{figure}[htbp]
  \begin{center}
  \includegraphics[height=0.7\textheight]{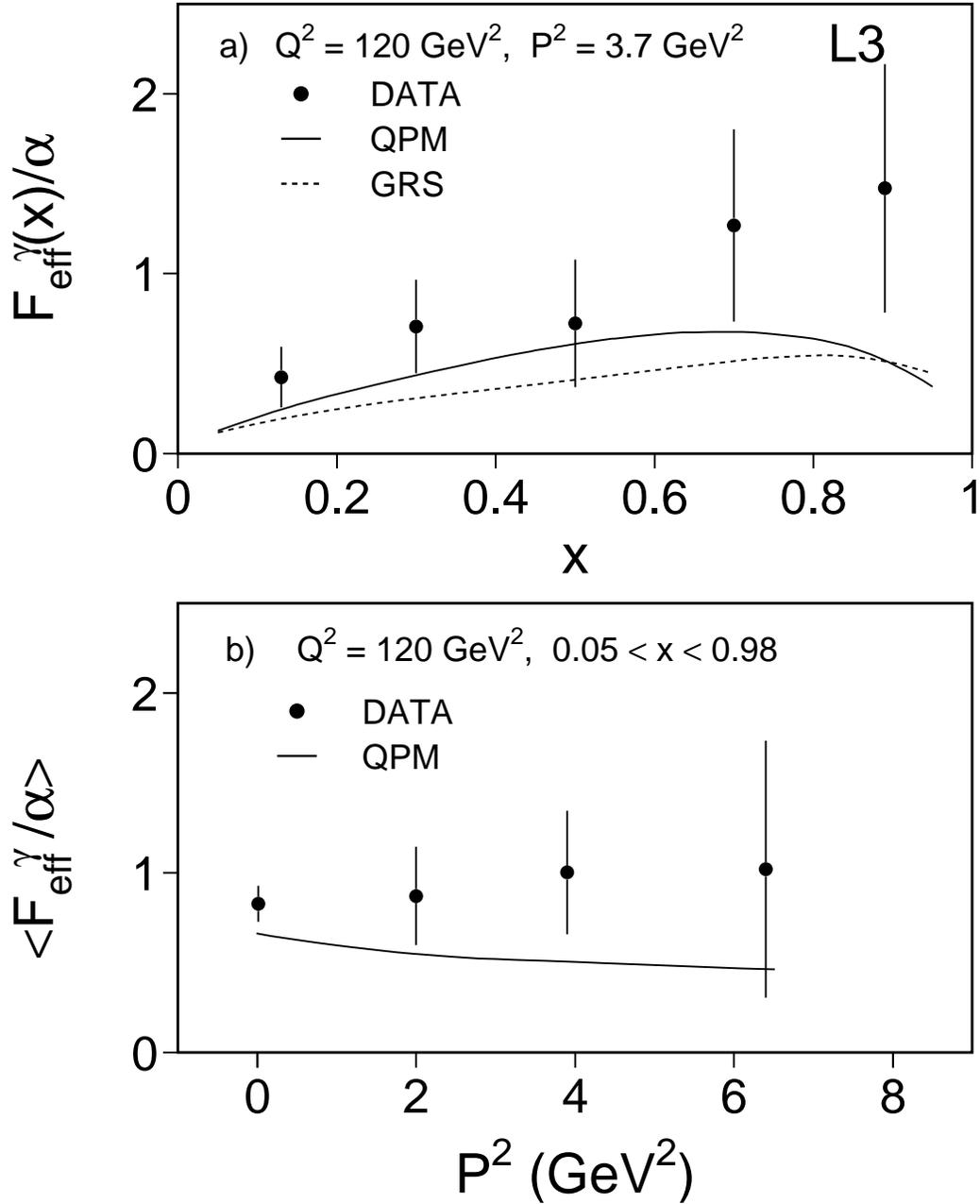}
  \caption{
      a) $\FE/\alpha$ for virtual photons at $ Q^2=120$ \GeV$^2$ and
      $P^2=3.7$ \GeV$^2$, compared with the QPM calculation 
      and the QCD calculation described in the text. 
      The QCD calculation considers only transverse photons;
      therefore it is not really comparable with the double-tag data.
      b) Dependence on $ P^2$ of $\FE/\alpha$ averaged over $x=0.05-0.98$ 
      for single-tag and double-tag data at $Q^2=120$ \GeV$^2$, compared
      with the QPM prediction, indicated as a full line.
      The errors are statistical and systematic added in quadrature.
      \label{fig:plot-fdouble}}
  \end{center}
\end{figure}

In Figure \ref{fig:plot-fdouble}a)
the data on $\FE/\alpha$ are compared with the QPM and the GRS
QCD model \cite{ref:GRS}.
%Gl\"uck, Reya and Stratmann (GRS)\cite{ref:GRS}.
The data are higher than the QPM prediction, but still compatible with it
within the combined statistical and systematic uncertainties.
The sensitivity of the QPM prediction to the c-quark mass, and at values
of $x$ larger than 0.9 to the u-quark mass, is similar to that for the
single-tag case. At lower values of $x$ the sensitivity to the value of the
u-quark mass is negligible.
A reduction of QCD effects has been predicted by Uematsu and Walsh
\cite{ref:Uematsu}, as the virtuality of the target photon limits
gluon emission.
In the present data their condition $ Q^2 \gg P^2 \gg \Lambda ^2_{\overline{MS}}$
is  fulfilled, as $ Q^2 = 120\hbox{ \GeV}^2$,
$ P^2 = 3.7\hbox{ \GeV}^2$ and $\Lambda ^2_{\overline{MS}}\approx 0.04$
\GeV$^2$. The GRS calculation pertains to the structure functions of
transverse photons only, neglecting completely the longitudinal photon
cross section; it is therefore impossible to draw a conclusion on the QCD
behaviour by comparing this calculation to the data.
\par
Figure \ref{fig:plot-fdouble}b) and Table 6 show the $ P^2$-dependence of
$\FE/\alpha$ averaged over the available $x$-range.
The average value  at $ P^2 = 3.7\hbox{ \GeV}^2$ is also listed in the Table.
The  data 
are above QPM expectations, but compatible  within the combined statistical
and systematic uncertainties.
%
%%%%%%%%%%%%%%%%%%%%%%%%%%%%%%%%%%%%%%%%%%%%%%%%%%%%%%%%%%%%%%%%%%%%%%%%%%%%%%%
% 
%  Conclusions
%
%%%%%%%%%%%%%%%%%%%%%%%%%%%%%%%%%%%%%%%%%%%%%%%%%%%%%%%%%%%%%%%%%%%%%%%%%%%%%%%
%

\section{Conclusions}
Two photon events $\rm e^+e^-\rightarrow e^+e^-  hadrons$ have been studied
in the high $Q ^2$ range $ 40 \GeV ^ 2 \le Q ^2 \le 500 \GeV ^ 2 $ 
at \rts $\simeq$ 91 \GeV {} in single-tag ($ P^2\approx 0 \GeV ^ 2$)
and double-tag ($ 1 \GeV ^ 2 \le P^2 \le 8 \GeV ^ 2$) mode.
\par
The  event shape distributions are well described by underlying point-like
interactions $ \gamma  \gamma  \rightarrow \qqbar $  (QPM),
$\rm \gamma g \rightarrow $\qqbar, $\rm \gamma  q \rightarrow g q$ (QCD),
whereas there is no evidence of a strong 
contribution of VDM and diffractive components.
The QPM diagram alone is insufficient to describe the observed $x$ distribution;
a QCD contribution at low values of $x$ improves the agreement with the data. 
 \par
The structure function $\FF(x)$ of real photons shows an excess at low-$x$ over
QPM and over several QCD calculations.
The observed $Q ^2$ evolution is compatible with the ln $Q^2$ dependence
measured previously \cite{ref:L3} at lower $Q ^2$ values.
 \par
An effective structure function is measured with double-tag events.
The value is higher than predicted by the QPM, but still compatible 
within the combined statistical and systematic uncertainties.
 \par

\section{Acknowledgments}
We express our gratitude to the CERN accelerator divisions for the excellent
performance of the LEP machine. We acknowledge with appreciation the effort
of the engineers, technicians and support staff who have participated in the
construction and maintenance of this experiment.
We wish to thank E.~Laenen, A.~Vogt and M.~Stratmann for useful discussions.
%
%%%%%%%%%%%%%%%%%%%%%%%%%%%%%%%%%%%%%%%%%%%%%%%%%%%%%%%%%%%%%%%%%%%%%%%%%%%%%%%
% The author list
%%%%%%%%%%%%%%%%%%%%%%%%%%%%%%%%%%%%%%%%%%%%%%%%%%%%%%%%%%%%%%%%%%%%%%%%%%%%%%%
%
\newpage
\section*{Author List}
\typeout{   }     
\typeout{Using author list for paper 204 -?}
\typeout{$Modified: Fri Feb  4 11:18:22 2000 by clare $}
\typeout{!!!!  This should only be used with document option a4p!!!!}
\typeout{   }
%
%
%
%  L A T E X  version!!
%
%
% Make sure that the Lep package has been used!
%\input{Lep.sty}%
%
%\ifx\LepCalled\undefined%
%\typeout{     }%
%\typeout{!!!!!!!!!!!!!!!!!!!!!!!!!!!!!!!!!!!!!!!!!!!!!!!!!!!!!!!!!!!}%
%\typeout{Yikes.  You haven't used the Lep package!}%
%\typeout{Please put \protect\usepackage\protect{Lep\protect} in your preamble,
%         followed by}%
%\typeout{\protect\Lep\protect{1\protect} or \protect\Lep\protect{2\protect}}%
%\typeout{     }%
%\typeout{For now you will get a Lep phase 2 authorlist (may not be right!).}%
%\typeout{!!!!!!!!!!!!!!!!!!!!!!!!!!!!!!!!!!!!!!!!!!!!!!!!!!!!!!!!!!!}%
%\typeout{     }%
%\Lep{2}\fi%

\newcount\tutecount  \tutecount=0
\def\tutenum#1{\global\advance\tutecount by 1 \xdef#1{\the\tutecount}}
\def\tute#1{$^{#1}$}
\tutenum\aachen            % 1
\tutenum\nikhef            % 2
\tutenum\mich              % 3
\tutenum\lapp              % 4
\tutenum\basel             % 5
\tutenum\lsu               % 6
\tutenum\beijing           % 7
\tutenum\berlin            % 8
\tutenum\bologna           % 9 
\tutenum\tata              % 10
\tutenum\ne                % 11
\tutenum\bucharest         % 12
\tutenum\budapest          % 13
\tutenum\mit               % 14 
\tutenum\debrecen          % 15
\tutenum\florence          % 16
\tutenum\cern              % 17 
\tutenum\wl                % 18 
\tutenum\geneva            % 19
\tutenum\hefei             % 20
\tutenum\seft              % 21
\tutenum\lausanne          % 22
\tutenum\lecce             % 23
\tutenum\lyon              % 24
\tutenum\madrid            % 25
\tutenum\milan             % 26
\tutenum\moscow            % 27
\tutenum\naples            % 27
\tutenum\cyprus            % 29
\tutenum\nymegen           % 30
\tutenum\caltech           % 31
\tutenum\perugia           % 32
\tutenum\cmu               % 33
\tutenum\prince            % 34
\tutenum\rome              % 35
\tutenum\peters            % 36
\tutenum\potenza           % 37
\tutenum\salerno           % 38
\tutenum\ucsd              % 39
\tutenum\santiago          % 40
\tutenum\sofia             % 41 
\tutenum\korea             % 42
\tutenum\alabama           % 43
\tutenum\utrecht           % 44
\tutenum\purdue            % 45
\tutenum\psinst            % 46
\tutenum\zeuthen           % 47
\tutenum\eth               % 48
\tutenum\hamburg           % 49
\tutenum\taiwan            % 50
\tutenum\tsinghua          % 51
{
\parskip=0pt
\noindent
{\bf The L3 Collaboration:}
\ifx\selectfont\undefined%  old style font selection
 \baselineskip=10.8pt
 \baselineskip\baselinestretch\baselineskip
 \normalbaselineskip\baselineskip
 \ixpt
\else%                      new style font selection
 \fontsize{9}{10.8pt}\selectfont
\fi
\medskip
\tolerance=10000
\hbadness=5000
\raggedright
\hsize=162truemm\hoffset=0mm
\def\r{\rlap,}
\noindent

M.Acciarri\r\tute\milan\
P.Achard\r\tute\geneva\ 
O.Adriani\r\tute{\florence}\ 
M.Aguilar-Benitez\r\tute\madrid\ 
J.Alcaraz\r\tute\madrid\ 
G.Alemanni\r\tute\lausanne\
J.Allaby\r\tute\cern\
A.Aloisio\r\tute\naples\ 
M.G.Alviggi\r\tute\naples\
G.Ambrosi\r\tute\geneva\
% ...
\newpage
%\rule{\textwidth}{0.4pt}
\begin{list}{A}{\itemsep=0pt plus 0pt minus 0pt\parsep=0pt plus 0pt minus 0pt
                \topsep=0pt plus 0pt minus 0pt}
\item[\aachen]
 I. Physikalisches Institut, RWTH, D-52056 Aachen, FRG$^{\S}$\\
 III. Physikalisches Institut, RWTH, D-52056 Aachen, FRG$^{\S}$
\item[\nikhef] National Institute for High Energy Physics, NIKHEF, 
     and University of Amsterdam, NL-1009 DB Amsterdam, The Netherlands
\item[\mich] University of Michigan, Ann Arbor, MI 48109, USA
\item[\lapp] Laboratoire d'Annecy-le-Vieux de Physique des Particules, 
     LAPP,IN2P3-CNRS, BP 110, F-74941 Annecy-le-Vieux CEDEX, France
\item[\basel] Institute of Physics, University of Basel, CH-4056 Basel,
     Switzerland
\item[\lsu] Louisiana State University, Baton Rouge, LA 70803, USA
\item[\beijing] Institute of High Energy Physics, IHEP, 
  100039 Beijing, China$^{\triangle}$ 
\item[\berlin] Humboldt University, D-10099 Berlin, FRG$^{\S}$
\item[\bologna] University of Bologna and INFN-Sezione di Bologna, 
     I-40126 Bologna, Italy
\item[\tata] Tata Institute of Fundamental Research, Bombay 400 005, India
\item[\ne] Northeastern University, Boston, MA 02115, USA
\item[\bucharest] Institute of Atomic Physics and University of Bucharest,
     R-76900 Bucharest, Romania
\item[\budapest] Central Research Institute for Physics of the 
     Hungarian Academy of Sciences, H-1525 Budapest 114, Hungary$^{\ddag}$
\item[\mit] Massachusetts Institute of Technology, Cambridge, MA 02139, USA
\item[\debrecen] KLTE-ATOMKI, H-4010 Debrecen, Hungary$^\P$
\item[\florence] INFN Sezione di Firenze and University of Florence, 
     I-50125 Florence, Italy
\item[\cern] European Laboratory for Particle Physics, CERN, 
     CH-1211 Geneva 23, Switzerland
\item[\wl] World Laboratory, FBLJA  Project, CH-1211 Geneva 23, Switzerland
\item[\geneva] University of Geneva, CH-1211 Geneva 4, Switzerland
\item[\hefei] Chinese University of Science and Technology, USTC,
      Hefei, Anhui 230 029, China$^{\triangle}$
\item[\seft] SEFT, Research Institute for High Energy Physics, P.O. Box 9,
      SF-00014 Helsinki, Finland
\item[\lausanne] University of Lausanne, CH-1015 Lausanne, Switzerland
\item[\lecce] INFN-Sezione di Lecce and Universit\'a Degli Studi di Lecce,
     I-73100 Lecce, Italy
\item[\lyon] Institut de Physique Nucl\'eaire de Lyon, 
     IN2P3-CNRS,Universit\'e Claude Bernard, 
     F-69622 Villeurbanne, France
\item[\madrid] Centro de Investigaciones Energ{\'e}ticas, 
     Medioambientales y Tecnolog{\'\i}cas, CIEMAT, E-28040 Madrid,
     Spain${\flat}$ 
\item[\milan] INFN-Sezione di Milano, I-20133 Milan, Italy
\item[\moscow] Institute of Theoretical and Experimental Physics, ITEP, 
     Moscow, Russia
\item[\naples] INFN-Sezione di Napoli and University of Naples, 
     I-80125 Naples, Italy
\item[\cyprus] Department of Natural Sciences, University of Cyprus,
     Nicosia, Cyprus
\item[\nymegen] University of Nijmegen and NIKHEF, 
     NL-6525 ED Nijmegen, The Netherlands
\item[\caltech] California Institute of Technology, Pasadena, CA 91125, USA
\item[\perugia] INFN-Sezione di Perugia and Universit\'a Degli 
     Studi di Perugia, I-06100 Perugia, Italy   
\item[\cmu] Carnegie Mellon University, Pittsburgh, PA 15213, USA
\item[\prince] Princeton University, Princeton, NJ 08544, USA
\item[\rome] INFN-Sezione di Roma and University of Rome, ``La Sapienza",
     I-00185 Rome, Italy
\item[\peters] Nuclear Physics Institute, St. Petersburg, Russia
\item[\potenza] INFN-Sezione di Napoli and University of Potenza, 
     I-85100 Potenza, Italy
\item[\salerno] University and INFN, Salerno, I-84100 Salerno, Italy
\item[\ucsd] University of California, San Diego, CA 92093, USA
\item[\santiago] Dept. de Fisica de Particulas Elementales, Univ. de Santiago,
     E-15706 Santiago de Compostela, Spain
\item[\sofia] Bulgarian Academy of Sciences, Central Lab.~of 
     Mechatronics and Instrumentation, BU-1113 Sofia, Bulgaria
\item[\korea]  Laboratory of High Energy Physics, 
     Kyungpook National University, 702-701 Taegu, Republic of Korea
\item[\alabama] University of Alabama, Tuscaloosa, AL 35486, USA
\item[\utrecht] Utrecht University and NIKHEF, NL-3584 CB Utrecht, 
     The Netherlands
\item[\purdue] Purdue University, West Lafayette, IN 47907, USA
\item[\psinst] Paul Scherrer Institut, PSI, CH-5232 Villigen, Switzerland
\item[\zeuthen] DESY, D-15738 Zeuthen, 
     FRG
\item[\eth] Eidgen\"ossische Technische Hochschule, ETH Z\"urich,
     CH-8093 Z\"urich, Switzerland
\item[\hamburg] University of Hamburg, D-22761 Hamburg, FRG
\item[\taiwan] National Central University, Chung-Li, Taiwan, China
\item[\tsinghua] Department of Physics, National Tsing Hua University,
      Taiwan, China
\item[\S]  Supported by the German Bundesministerium 
        f\"ur Bildung, Wissenschaft, Forschung und Technologie
\item[\ddag] Supported by the Hungarian OTKA fund under contract
numbers T019181, F023259 and T024011.
\item[\P] Also supported by the Hungarian OTKA fund under contract
  numbers T22238 and T026178.
\item[$\flat$] Supported also by the Comisi\'on Interministerial de Ciencia y 
        Tecnolog{\'\i}a.
\item[$\sharp$] Also supported by CONICET and Universidad Nacional de La Plata,
        CC 67, 1900 La Plata, Argentina.
\item[$\diamondsuit$] Also supported by Panjab University, Chandigarh-160014, 
        India.
\item[$\triangle$] Supported by the National Natural Science
  Foundation of China.
\item[\dag] Deceased.
\end{list}
}
\vfill

%%% Local Variables: 
%%% mode: latex
%%% TeX-master: t
%%% End:

\newpage
%
%
%%%%%%%%%%%%%%%%%%%%%%%%%%%%%%%%%%%%%%%%%%%%%%%%%%%%%%%%%%%%%%%%%%%%%%%%%%%%%%%
% 
%  Bibliography
%
%%%%%%%%%%%%%%%%%%%%%%%%%%%%%%%%%%%%%%%%%%%%%%%%%%%%%%%%%%%%%%%%%%%%%%%%%%%%%%%
%

\pagebreak

%
%%%%%%%%%%%%%%%%%%%%%%%%%%%%%%%%%%%%%%%%%%%%%%%%%%%%%%%%%%%%%%%%%%%%%%%%%%%%%%%
% 
%  Tables
%
%%%%%%%%%%%%%%%%%%%%%%%%%%%%%%%%%%%%%%%%%%%%%%%%%%%%%%%%%%%%%%%%%%%%%%%%%%%%%%%
%

{\small
\begin{center}
\vspace{0.2 cm}
\begin{tabular}{|l|c|c|c|c|c|c|c|c|c}
\hline
                                       & single tag events & double tag events \\
\hline
Data                                     & 496         & 43   \\
\hline
Background  &   &  \\
$ \rm e^+e^- \ra e^+e^-\tau^+\tau^-$        &$24\pm 6$    & $<1$   \\
$\rm e^+e^-  \ra q\bar q$                   &$12\pm 7$    & $<2$   \\
$\rm e^+e^-  \ra \tau^+\tau^-$              &$\phantom{0} 4\pm 2$    & $<1$   \\
\hline
Data $-$ Background                       &$456\pm 24$  &$43\pm 7$\\
\hline
JAMVG$\phantom{00000}$                   &            &           \\
$\rm \phantom{00000}u\bar u,d\bar d,s\bar s$ &$233\pm 2$  &$15\pm 1$  \\
$\rm \phantom{00000}c\bar c$                 &$111\pm 1$  &$10\pm 0$  \\
Total$\phantom{00000}$                   &$344\pm 2$  &$25\pm 1$  \\
\hline
PHOJET                                   &$346\pm 2$  &$48\pm 1$  \\
\hline
TWOGAM                                   &            &            \\
$\phantom{00000}$ QPM                    &$335\pm 5$  &$24\pm 1$   \\
$\phantom{00000}$ VDM                    &$126\pm 3$  &$\phantom{0}9\pm 1$    \\
$\phantom{00000}$ QCD                    &$162\pm 1$  &$29\pm 1$   \\
Total$\phantom{00000}$                   &$624\pm 6$  &$62\pm 2$   \\
\hline
\end{tabular}
\end{center}
\begin{center}
Table 1: Numbers of selected events and Monte Carlo predictions normalized to the
luminosity of the data.
\end{center}
}

{\small
\begin{center}
\vspace{0.2 cm}
\begin{tabular}{|c|r|r|r|r|r|r|r|r|r}
\hline
     $x$ range            & $0.05-0.2$  & $0.2-0.4$ & $0.4-0.6$ & $0.6-0.8$ & $0.8-0.98$\\
$\langle x\rangle $       & 0.13    & 0.30  & 0.50  & 0.70  & 0.89  \\
\hline
 Ratio to QPM             & $2.26\pm 0.27$ & $1.56\pm 0.16$ & $1.03\pm 0.16$
                          & $0.95\pm 0.15$  & $1.18\pm 0.25$  \\
\hline
       $\FF(x)/\alpha$    & $0.66\pm 0.08$ & $0.81\pm 0.08$ & $0.76\pm 0.12$
                          & $0.85\pm 0.14$ & $0.91\pm 0.19$\\
Systematic uncertainty          & 0.06    & 0.08  & 0.07  & 0.08  & 0.09 \\
\hline
\hbox{  Correlation}& 1  & 0.18 & $-0.19$ & $-0.02$ &  0.02 \\
\hbox{  matrix}     &    & 1    &  0.03 & $-0.18$ &  0.00 \\
                    &    &      &  1    &  0.09 & $-0.21$ \\
                    &    &      &       &  1    &  0.04 \\
                    &    &      &       &       &  1    \\
\hline
\end{tabular}
\end{center}
\begin{center}
Table 2: $\FF/\alpha$ as a function of $x$ for real photons at
$Q^2=120$ \GeV$^2$, from single-tag events.
\end{center}
}

{\small
\begin{center}
\vspace{0.2 cm}
\begin{tabular}{|l|c|c|}
\hline
 Source of systematic uncertainty                  & Single tag & Double tag \\
\hline
 Monte Carlo statistics                     & 1.7\%      & 5.8\%      \\
 Tag selection                              & 1.0\%      & 1.4\%      \\
 Model dependence                           & 4.6\%      & 4.6\%      \\
 Triggering                                 & 1.0\%      & 1.0\%      \\
 Event selection                            & 3.0\%      & 1.6\%      \\
 Radiative corrections                      & 3.2\%      & 3.2\%      \\
 Modelling of background                     & 4.4\%      & 4.7\%      \\
 Unfolding                                  & 5.3\%      & 8.1\%      \\
\hline
                               Total        & 9.6\%      & 12.6\%      \\
\hline
\end{tabular}
\end{center}
\begin{center}
Table 3: Systematic relative uncertainties on $\FF(x,Q^2)/\alpha$ and
$\FE(x,Q^2,P^2)/\alpha$.
\end{center}
}
\pagebreak

{\small
\begin{center}
\vspace{0.2 cm}
\begin{tabular}{|c|c|c|c|}
\hline
$\langle Q^2\rangle$&$\langle \FF/\alpha\rangle$ & $\langle \FF/\alpha\rangle$
&$\langle\FF/\alpha\rangle$  \\
\GeV$^2$   &     $x=0.05-0.98$       &      $x=0.3-0.8$       &$x=0.1-0.6$ \\
\hline
 60     & $0.73\pm 0.11\pm 0.07$ &$0.66\pm 0.09\pm 0.06$&$0.63\pm 0.06\pm 0.06$
                                               \\
 90     & $0.89\pm 0.13\pm 0.09$ &$0.79\pm 0.14\pm 0.08$&$0.92\pm 0.14\pm 0.09$
                                              \\
 125    & $0.85\pm 0.11\pm 0.09$ &$0.88\pm 0.12\pm 0.08$&$0.86\pm 0.14\pm 0.08$
                                              \\
 225    & $1.01\pm 0.25\pm 0.10$ &$1.18\pm 0.22\pm 0.11$&$0.91\pm 0.30\pm 0.09$
                                              \\
\hline
 120    & $0.83\pm 0.06\pm 0.08$ &$0.78\pm 0.06\pm 0.08$&$0.71\pm 0.05\pm 0.07$
                                              \\
\hline
\end{tabular}
\end{center}
\begin{center}
Table 4:  $Q^2$ dependence of $\FF/\alpha$ for real photons for various
$x$ intervals. In the last row the structure function value is given 
for the total sample of single-tag events.
\end{center}
}

{\small
\begin{center}
\vspace{0.2 cm}
\begin{tabular}{|c|r|r|r|r|r|r|r|r|r}
\hline
     $x$ range            & $0.05-0.2$  & $0.2-0.4$ & $0.4-0.6$ & $0.6-0.8$ & $0.8-0.98$\\
     $\langle x\rangle $                  & 0.13    & 0.3   & 0.5   & 0.7   & 0.89 \\
\hline
 Ratio to QPM             & $1.85\pm 0.70$ & $1.63\pm 0.56$ & $1.19\pm 0.57$ 
                          & $1.89\pm 0.76$ & $3.01\pm 1.33$ \\
\hline
          $\FE(x)/\alpha$ & $0.42\pm 0.16$ & $0.71\pm 0.24$ & $0.72\pm 0.34$
                          & $1.27\pm 0.51$ & $1.48\pm 0.66$ \\
Systematic uncertainty          & 0.05    & 0.09  & 0.09  & 0.16  & 0.19 \\
\hline
\hbox{ Correlation} & 1 & $-0.02$& $-0.08$ & 0.01  & 0.01  \\
\hbox{ matrix}      &   & 1    &  0.09 & $-0.12$ & $-0.04$ \\
                    &   &      &  1    &  0.12 &  0.13 \\
                    &   &      &       & 1     &  0.50 \\
                    &   &      &       &       &  1    \\
\hline
\end{tabular}
\end{center}
\begin{center}
Table 5:  $\FE/\alpha$ as function of $x$ for virtual photons at
$P^2=3.7$ \GeV$^2$, $Q^2=120$ \GeV$^2$, from double-tag events.
\end{center}
}
{\small
\begin{center}
\vspace{0.2 cm}
\begin{tabular}{|c|c|c|}
\hline
 $\langle P^2\rangle$ &$\langle \FE/\alpha\rangle$     \\
 \GeV$^2$ & $x=0.05-0.98$   \\
\hline
 0  &$0.83\pm 0.06\pm 0.08$\\
 2.0&$0.87\pm 0.25\pm 0.11$\\
 3.9&$1.00\pm 0.32\pm 0.13$\\
 6.4&$1.02\pm 0.70\pm 0.13$\\
\hline
3.7&$0.94\pm 0.19\pm 0.12$\\
\hline
\end{tabular}
\end{center}
\begin{center}
Table 6: 
 $P^2$ dependence of $\FE/\alpha$ for virtual photons
at $Q^2=120$ \GeV$^2$. In the last row the structure function value is given for the 
total sample of double-tag events.
\end{center}
}
\vfill
\pagebreak

%
%%%%%%%%%%%%%%%%%%%%%%%%%%%%%%%%%%%%%%%%%%%%%%%%%%%%%%%%%%%%%%%%%%%%%%%%%%%%%%%
% 
%  Figures
%
%%%%%%%%%%%%%%%%%%%%%%%%%%%%%%%%%%%%%%%%%%%%%%%%%%%%%%%%%%%%%%%%%%%%%%%%%%%%%%%
%

\end{document}